\documentclass{aa}  

\newcommand{\hrieuv}{\text{HRI}_{\text{EUV}}}
\usepackage{graphicx}
\usepackage{bm}
\usepackage{natbib}
\usepackage{mathtools}
\usepackage{amsmath}
\usepackage{multirow}
\usepackage[export]{adjustbox}
\usepackage{rotating}
\usepackage{tikz}
\usepackage{txfonts}
\begin{document}

   \title{Characteristics and Energy Flux Distributions of Decayless Transverse Oscillations Depending on Coronal Regions}
   \subtitle{Decayless Transverse Oscillations Depending on Coronal Regions}

   \author{%
            {Daye Lim,}\inst{\ref{aff:ROB}, \ref{aff:CmPA}}
            {Tom Van Doorsselaere,}\inst{\ref{aff:CmPA}}
            {David Berghmans}\inst{\ref{aff:ROB}}
            \and {Elena Petrova}\inst{\ref{aff:CmPA}}
          }

   \institute{%
            \label{aff:ROB}{Solar-Terrestrial Centre of Excellence – SIDC, Royal Observatory of Belgium, Ringlaan -3- Av. Circulaire, 1180 Brussels, Belgium}\\ \email{daye.lim@oma.be}
            \and
            \label{aff:CmPA}{Centre for mathematical Plasma Astrophysics, Department of Mathematics, KU Leuven, Celestijnenlaan 200B, 3001 Leuven, Belgium}
             }

   \authorrunning{Lim et al.}


  \abstract
   {\citet{2023ApJ...952L..15L} have recently proposed that the slope ($\delta$) of the power law distribution between the energy flux and oscillation frequency could determine whether high-frequency transverse oscillations give a dominant contribution to the heating ($\delta<1$). A meta-analysis of decayless transverse oscillations revealed that high-frequency oscillations potentially play a key role in heating the solar corona.}
   {We aim to investigate how (whether) the distributions of the energy flux contained in transverse oscillations, and their slopes, depend on the coronal region in which the oscillation occurs.}
   {We analyse transverse oscillations from 41 quiet Sun (QS) loops and 22 active region (AR) loops observed by Solar Orbiter/Extreme Ultraviolet Imager (SolO/EUI) $\hrieuv$. The energy flux and energy are estimated using analysed oscillation parameters and loop properties, such as periods, displacement amplitudes, loop lengths, and minor radii of the loops.}
   {It is found that about 71\% of QS loops and 86 \% of AR loops show decayless oscillations. We find that the amplitude does not change depending on different regions, but the difference in the period is more pronounced. Although the power law slope ($\delta=-1.79$) in AR is steeper than that ($\delta=-1.59$) in QS, both of them are significantly less than the critical slope of 1.}
   {Our statistical study demonstrates that high-frequency transverse oscillations can heat the QS. For ARs, the total energy flux is insufficient unless yet-unobserved oscillations with frequencies up to 0.17 Hz are present. Future EUI campaigns will be planned to confirm this.}


   \maketitle

\section{Introduction}

Space-based imaging observations with high spatial resolution, Solar Dynamics Observatory/Atmospheric Imaging Assembly (SDO/AIA; \citealt{2012SoPh..275...17L}) and Solar Orbiter/Extreme Ultraviolet Imager (SolO/EUI; \citealt{2020A&A...642A...8R}), have significantly shaped queries on decayless transverse (or kink) oscillations in recent years (see \citealt{2021SSRv..217...73N}, for a recent review). This oscillation shows transverse motions of the coronal loop axis with a small amplitude \citep{2012ApJ...759..144T, 2022MNRAS.516.5989Z, 2023NatSR..1312963Z, 2023NatCo..14.5298Z}. This mode can be observed without significant damping for several oscillation cycles \citep{2022MNRAS.513.1834Z}. Thus, this oscillation has been termed decayless to distinguish it from the standard transverse oscillations, excited by solar energetic events, that show rapid damping \citep{2013A&A...552A..57N}. 

Prior research confirms that decayless oscillations are a common feature in the solar corona and can exist without a noticeable energetic event \citep{2013A&A...560A.107A}. \citet{2015A&A...583A.136A} found that 90\% of 21 active regions (ARs) detected with SDO/AIA showed decayless oscillations. It was observed that among 23 coronal bright points in the quiet Sun (QS) observed by SDO/AIA, decayless oscillations occurred in 70\% of them \citep{2022ApJ...930...55G}. Recently, \citet{2024A&A...685A..36S} confirmed that using SolO/EUI, closed loops in coronal holes (CHs) can exhibit decayless signatures. However, \citet{2022A&A...666L...2M} detected that an identical loop system in an AR does not always display decayless oscillations, although the physical conditions remain approximately the same. 

Decayless oscillations have been observed in loops with lengths between about 3 and 740 Mm, and the observed periods were between about 11 s and 30 min \citep{2023ApJ...944....8L, 2023NatSR..1312963Z, 2024A&A...685A..36S}. The relationship between loop lengths and periods has shown two tendencies. One relationship is linear, implying that decayless oscillations could be interpreted as standing modes \citep{2015A&A...583A.136A, 2023ApJ...944....8L}. Furthermore, \citet{2018ApJ...854L...5D} found an observation of the co-existence of the fundamental and second harmonics of decayless oscillations, which can support the interpretation as standing modes. The other relationship is no correlation \citep{2022ApJ...930...55G, 2024A&A...685A..36S}. To identify the mode of these oscillations, there were additional efforts to investigate phase relations along the loop axis, but no significant propagating features were found \citep{2023ApJ...946...36P, 2024A&A...685A..36S}. Hence, the interpretation of these decayless oscillations remains open.

Even after a considerable number of decayless oscillations observed, it remains unsolved how decayless oscillations are sustained against damping. Considering the rapid damping mechanism of large-amplitude transverse oscillations to be an inherent feature \citep{2002A&A...394L..39G, 2008ApJ...687L.115T}, it can be expected that there is a continuous energy input from the convection zone to keep decayless oscillations lasting for several cycles. Numerical studies have shown that decayless oscillations could be excited by a p-mode driver \citep{2023ApJ...955...73G}, harmonic driver \citep{2017A&A...604A.130K, 2019ApJ...870...55G}, supergranulation flows \citep{2016A&A...591L...5N, 2020ApJ...897L..35K}, and broadband drivers \citep{2020A&A...633L...8A, 2024A&A...681L...6K}. However, there have also been claims that decayless oscillations are not actual kink oscillations of the full loop body. A numerical simulation found that decayless features can result from the combination of periodic brightenings and the motions produced by the Kelvin–Helmholtz instabilities (KHi) \citep{2016ApJ...830L..22A}. Recently, \citet{2024MNRAS.527.5741L} reported theoretically that radial motion caused by slow magnetoacoustic oscillations in a short loop can appear as decayless transverse oscillations.

Although there are still uncertainties about the excitation and dissipation of decayless oscillations, their total energy content would be of great interest given their ubiquity. Recently, \citet{2023ApJ...952L..15L} have introduced the wave-based heating theory similar to the nanoflare heating theory \citep{1991SoPh..133..357H}. If transverse oscillations of varying frequencies generate energy flux, then the total energy flux, $F$, is equal to the integral of spectral energy flux, $s(\omega)$, which is the energy flux per frequency,
\begin{equation}\label{eq:totalenergyflux_freq}
F=\int_{\omega_{\text{min}}}^{\omega_{\text{max}}}s(\omega)d\omega,
\end{equation}
where $s(\omega)$ has dimensions of $\text{W}\,\text{m}^{-2}\,\text{Hz}^{-1}$ and the limits $\omega_{\text{min}}$ and $\omega_{\text{max}}$ are the lowest and highest frequencies of transverse oscillations, respectively. If the spectral energy flux, $s(\omega)$, follows a power law of the form,
\begin{equation}\label{eq:powerlaw}
s(\omega)=s_{0}\omega^{-\delta},
\end{equation}
where $s_{0}$ is a scaling constant and $\delta$ is a power-law slope. Then, assuming that $\delta < 1$, 
\begin{equation}\label{eq:approx_energyflux_high}
F_{\omega}\approx\frac{s_{0}}{-\delta+1}\omega_{\text{max}}^{-\delta+1},
\end{equation}
implying that high-frequency transverse oscillations provide the dominant contribution to the heating generated by transverse oscillations. If, however, $\delta > 1$, then
\begin{equation}\label{eq:approx_energyflux_low}
F_{\omega}\approx\frac{s_{0}}{\delta-1}\omega_{\text{min}}^{-\delta+1},
\end{equation}
and low-frequency transverse oscillations dominate the heating. Through a meta-analysis using the literature reporting decayless transverse oscillations, \citet{2023ApJ...952L..15L} found $\delta$ of around -1.4 for decayless oscillations in the frequency range $0.002-0.07$ Hz (corresponding to 14-500 s). This result indicated that high-frequency decayless oscillations could give the dominant heating in the corona, compared to low-frequency oscillations. 

However, considering that different coronal regions have different magnetic topologies and energy losses in coronal regions, this analysis should be repeated per region. \citet{2016ApJ...828...89M} found that the spectral slope of propagating transverse waves observed by the Coronal Multi-channel Polarimeter varied between active regions, quiet Sun, and open field regions. In this paper, we study the statistical characteristics of decayless transverse oscillations in different regions using SolO/EUI observations, in order to investigate the energy flux distribution and the power law slope in each region. Based on the theory proposed in \citet{2023ApJ...952L..15L}, the highest frequency ($\omega_{\text{max}}$) of decayless transverse oscillations is a key parameter to estimate the total energy flux generated by all decayless oscillations. Thus, it would be worthwhile to further explore what the highest observable frequency of transverse oscillations is by using the high spatial and temporal resolutions of SolO/EUI. In Section \ref{sec:data} we describe our data sets. The results and discussion are featured in Section \ref{sec:result}. Section \ref{sec:conclusion} gives conclusions.

\section{Data and Analysis}\label{sec:data}

\subsection{Data}
We consider the calibrated level-2 SolO/EUI High Resolution Imager 174 \AA\,($\hrieuv$) data \citep{euidatarelease6}. Among the released data, only data sequences with a cadence of equal to or less than 3 seconds are used to expand the limit of observable high frequencies of oscillations. Note that we do not limit the upper range of periods with this criteria. We exclude observations that were already used in the study of decayless transverse oscillations \citep{2023ApJ...944....8L, 2023ApJ...946...36P, 2024A&A...685A..36S} or observations that were saturated and/or belonged to technical commissioning tests. As a consequence, we use 16 data sequences with varying spatial scales observing both QS and ARs. The details of the datasets are listed in Table \ref{tab:data}. Note that there were 3 data sequences observing CHs with a high cadence. One was used in \citet{2024A&A...685A..36S} and the others were saturated. Thus, it was not possible to include CH observations in this study. 

We remove the jitter of the spacecraft in the level-2 data using a cross-correlation technique \citep{2022A&A...667A.166C}. Although this technique has a sub-pixel accuracy, given that the decayless oscillations have very low amplitudes, we additionally examine the periodicity of the remaining jitter after applying the technique. We consider wavelet analysis to the remaining jitter and find that none of the 16 data sequences had any significant periodicity.  

\begin{table*}[ht]
\caption{Details of the datasets used in this study.}
\label{tab:data}
\begin{tabular}{ccccccc}
\hline
Date       & Start Time (UT) & End Time (UT) & $D_{\text{Sun}}$ (au) & Pixel scale (km) & Cadence (s) & Region                     \\ \hline
\hline
2020-05-21 & 19:02:30    & 19:03:29  & 0.60        & 215              & 3           & Quiet Sun                  \\
2020-05-21 & 19:03:29    & 19:04:08  & 0.60        & 215              & 2           & Quiet Sun                  \\
2020-05-21 & 20:00:11    & 20:00:28  & 0.60        & 215              & 1           & Quiet Sun                  \\
2021-02-19 & 23:50:40    & 23:51:00  & 0.51        & 183              & 2           & Quiet Sun                  \\
2021-09-13 & 00:11:51    & 00:24:59  & 0.59        & 212              & 2           & Quiet Sun                  \\
2022-03-08 & 00:00:03    & 00:29:21  & 0.49        & 176              & 3           & Quiet Sun                  \\
2022-03-17 & 00:18:00    & 00:47:57  & 0.38        & 136              & 3           & Quiet Sun                  \\
2022-03-30 & 00:03:00    & 00:47:57  & 0.33        & 118              & 3           & Active Region (NOAA 12974) \\
2022-10-12 & 05:25:00    & 06:09:27  & 0.29        & 104              & 3           & Quiet Sun                  \\
2022-10-13 & 13:06:00    & 13:58:18  & 0.29        & 104              & 3           & Active Region (Farside)    \\
2023-03-29 & 12:40:00    & 13:30:00  & 0.39        & 140              & 3           & Active Region (NOAA 13262) \\
2023-04-04 & 06:18:08    & 06:47:59  & 0.33        & 118              & 3           & Quiet Sun                  \\
2023-04-07 & 04:20:00    & 05:10:00  & 0.30        & 108              & 3           & Active Region (NOAA 13270) \\
2023-04-10 & 20:59:55    & 21:49:55  & 0.29        & 104              & 3           & Quiet Sun                  \\
2023-04-11 & 23:14:55    & 23:44:52  & 0.29        & 104              & 3           & Quiet Sun                  \\
2023-04-15 & 04:43:00    & 05:17:21  & 0.31        & 111              & 3           & Quiet Sun                  \\ \hline
\end{tabular}
\tablefoot{The region for each data sequence is classified by comparing $\hrieuv$ observation with SDO/AIA and Helioseismic and Magnetic Imager (HMI). When the SolO is located farside and the co-observation with the SDO is unable, the classification is decided by eye.}
\end{table*}

\subsection{Analysis}

\begin{figure}
  \resizebox{\hsize}{!}{\includegraphics{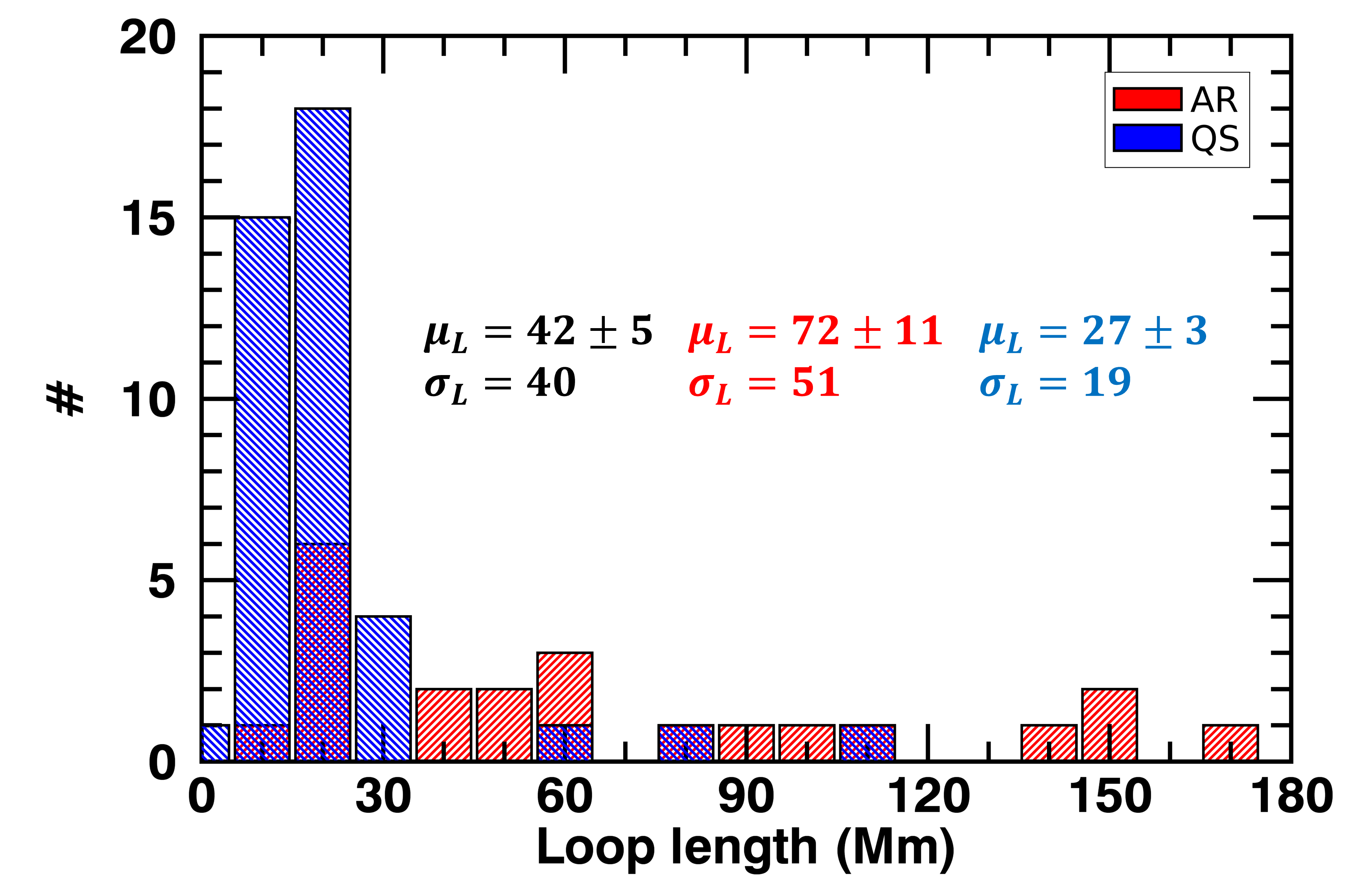}}
  \caption{Histogram of the lengths of the loops in quiet Sun (blue hatched bar) and active regions (red hatched bar). The average and standard deviation of loop lengths for quiet Sun (blue), active regions (red), and all regions (black) are indicated in the panel.}
  \label{fig:hist_loop}
\end{figure}

\begin{figure*}
  \resizebox{\hsize}{!}{\includegraphics{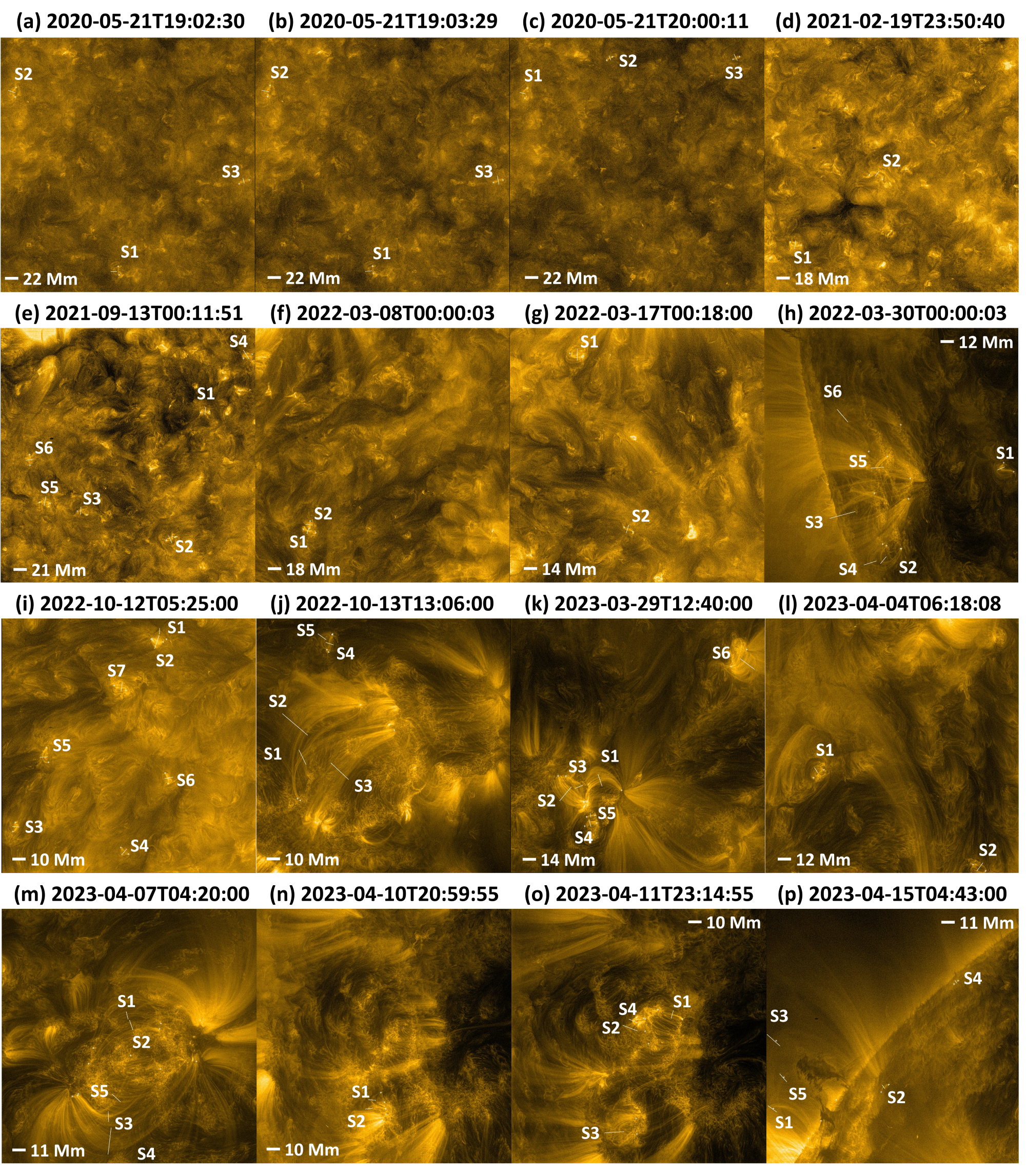}}
  \caption{SolO/EUI HRIEUV 174 $\AA$ full FOV images for each of the 16 data sets. The image has been enhanced using the Multiscale Gaussian Normalization \citep{2014SoPh..289.2945M}. The observation date and start time are displayed above each image. In each panel, the white solid lines mark the positions of the artificial slits that are used for generating the time-distance maps. The white pluses in each panel show the approximate position of footpoints for on-disc loops and the position of the apex and the center between two footpoints for limb loops. Some loops share the same footpoint. The details of the data set are provided in Table \ref{tab:data}. The spatial scale is indicated as a white thick bar in each panel.}
  \label{fig:datasets}
\end{figure*}

We analyse transverse oscillations of 41 loops in QS and 22 loops in ARs, ranging from small to large scales. Using EUV images only, we consider closed loops that can be identified by the eye contrasted to the background and whose footpoints are relatively stationary throughout the observation duration. However, since the loop itself and its surroundings are very dynamic, some loops that do not maintain the brightness throughout their duration are included. The loop length ($L$) is approximated by measuring the distance ($D_{f}$) between footpoints and using the relation, $L=\pi D_{f}/2$, assuming that the loop is a semi-circle. In the case of the loops on the limb, $D_{f}$ is measured by considering the distance between the loop apex and the center between two footpoints. The estimated loop lengths range from about 7 to 174 Mm (Figure \ref{fig:hist_loop}). The average loop length in AR is around 70 Mm, which is longer than that of around 30 Mm in QS. This is shorter than the length of AR loops mainly observed in AIA for decayless oscillations (an average of about 220 Mm in \citealt{2015A&A...583A.136A}) and includes much longer loops than the range (about 3-30 Mm) of loop lengths considered for decayless oscillations in EUI \citep{2023ApJ...944....8L, 2023ApJ...946...36P, 2024A&A...685A..36S}. We consider time-series data from an artificial slit on the loop apex (see Figure \ref{fig:datasets} for slit positions in each data sequence). To improve the signal-to-noise ratio, we consider a slit of thickness 5 pixels and average the intensity over the slit thickness.

Considering the low amplitude feature of decayless oscillations and the large data samples considered in this study, tracking the transverse motions of each loop manually is arduous. To facilitate the analysis, we employ the Automatic Northumbria University Wave Tracking\footnote{\url{https://github.com/Richardjmorton/auto_nuwt_public}} (Auto-NUWT; \citealt{2013ApJ...768...17M, 2018ApJ...852...57W}) code to detect the position of the loops, track their transverse motion, and extract the properties of their oscillations. To track the loop position, this code finds the position where local intensity gradients in both transversal directions are higher than a user-adjustable threshold. A constant threshold of 0.5 for unmasked time-distance maps was used in studies mostly targeting off-limb regions \citep{2014ApJ...790L...2T, 2020ApJ...894...79W}. In contrast, the regions considered in this study are mostly observed on disc. High-resolution observations sharpen not only the coronal loop itself but also features in the loop background. For this reason, the intensity threshold for defining a loop is expected to be highly dependent on regions. We consider the average gradient of each time-distance map as the threshold for each map. 

\begin{figure}
  \resizebox{\hsize}{!}{\includegraphics{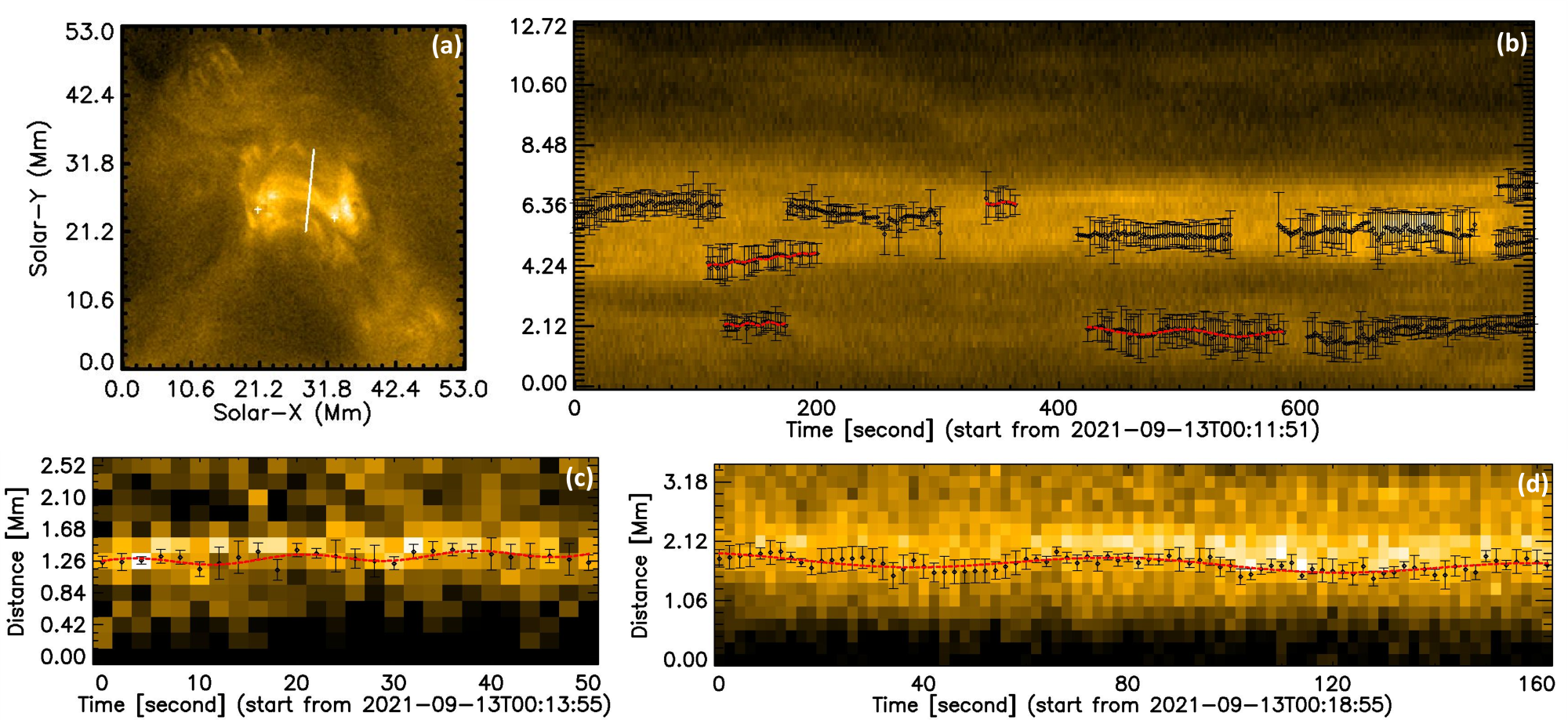}}
  \caption{(a) A magnified view of the loop with slit S1 (white solid line) and its footpoints (white plus) on 2021-09-13. (b) Time-distance map made from the slit presented in the panel (a). The black diamond and error bars indicate the detected loop centers and minor radius. The red dashed line outlines their sine fits that meet all criteria. The close-up view of fitted oscillations is shown in (c) and (d). In this case, error bars represent position errors.}
  \label{fig:tdmapexample}
\end{figure}

Based on the local intensity maxima filtered out by the threshold, the fitting of the nearby intensity values with a Gaussian function is considered. From this, the center and minor radius of the loop are detected as the Gaussian center and half width at half maximum respectively (see Figure \ref{fig:tdmapexample}). The fitting is weighted by $\hrieuv$ intensity errors ($\sigma_{I}$) which are taken as
\begin{equation}\label{eq:intensityerror}
\sigma_{I}\approx\sqrt{r^2+\frac{It_{\text{exp}}\alpha}{N}},
\end{equation}
where $r=2\,\text{DN}$ is the detector readout noise, $I$ is the intensity, $t_{\text{exp}}$ is the exposure time, $\alpha=6.85\,\text{DN}\,\text{photon}^{-1}$ is the photons to DN conversion factor, and $N$ is the number of pixels over which the intensity is averaged (i.e., 5 in this study) \citep{2023ApJ...946...36P, 2023Sci...381..867C, 2023arXiv230714182G}. The number of transverse pixels used in the fit could affect the convergence. We considered six different pixel sizes of 11, 13, 15, 17, 19, and 21, corresponding to the transverse scales between around 0.5 and 2 Mm. The pixel size of 13 was chosen as it matched best with the loops seen in the time-distance maps. 

Transverse oscillatory features of the loop center are identified using a Fourier analysis with a 95\% significance level. In order to obtain only robust transverse oscillations against potential false detected ones, we accept oscillations where they meet the following criteria.
\begin{enumerate}
    \item The fit should be made using more than five points \citep{2014ApJ...790L...2T}. This directly gives the minimum period of oscillations, 5 s from the 1-s cadence data sequences, 10 s from the 2-s cadence, and 15 s from the 3-s cadence.
    \item The oscillation duration should be longer than one period.
    \item The percentage of time frames in which the loop center is not detected during the total duration of the oscillation should be less than 35\% \citep{2018ApJ...852...57W}.   
\end{enumerate}
Our procedure identifies 412 single-period oscillations and 178 multi-period oscillations. The multi-period oscillations may represent a possibility of multiple harmonics as they were already discovered observationally \citep{2018ApJ...854L...5D} and numerically \citep{2021SoPh..296..124R, 2024A&A...681L...6K}. To identify them, however, the phase relation along the loop axis of each oscillation should be carefully analyzed, which is beyond the scope of this study. In this study, we only consider oscillations that have a single dominant period and can be well described as a sinusoidal function, which implies the absence of damping and is mainly used in analysing decayless oscillations in previous studies \citep{2023ApJ...944....8L, 2023ApJ...946...36P, 2024A&A...685A..36S}. Considering the oscillation parameters as initial guesses, we fit the data for 412 cases with a sine and a linear trend
\begin{equation}
    X(t)=A\sin{\left(\frac{2\pi t}{P}+\phi\right)}+X_0+X_1t,
\end{equation}
where $X$ is the transverse displacement at $t$, $A$ is the displacement amplitude, $P$ is the period, $\phi$ is the phase, and $X_0$ and $X_1$ are the constant parameters of the linear trend. In order to examine the goodness of fit and filter out the fits strictly, we consider two criteria as follows.
\begin{enumerate}
    \item We accept a fit if the error ($\delta A$ and $\delta P$) of fitted parameters ($A$ and $P$) are not comparable to them: $\sqrt{(\delta A/A)^2+(\delta P/P)^2}\leq0.7$ \citep{2014ApJ...790L...2T}.
    \item We accept a fit if the chi-squared is less than three times the number of data points \citep{2009ASPC..411..251M}.   
\end{enumerate}
After applying these criteria, 254 oscillations are retained out of the original 412, and their parameters and loop information are summarised in Table \ref{tab:parameters}. An example of the fitted oscillations is shown in Figure \ref{fig:tdmapexample}.



\begin{table*}[]
\caption{Parameters of 254 decayless transverse oscillations of coronal loops detected with SolO/EUI $\hrieuv$.}
\label{tab:parameters}
\begin{adjustbox}{width=1\textwidth}
\small
\begin{tabular}{cccccccccccc}
\hline
Date       & Start time & End time & Slit ID & Loop length (Mm) & Displacement amplitude (km) & Amplitude uncertainty (km) & Period (s) & Period uncertainty (s) & Duration (s) & Minor radius (km) & Intensity ratio \\ \hline \hline
2020-05-21 & 19:02:30   & 19:03:15 & 1       & 23               & 100                         & 41                         & 31.2       & 2.5                    & 48           & 216               & 1.30            \\
2020-05-21 & 19:02:30   & 19:03:27 & 1       & 23               & 63                          & 27                         & 18.3       & 1.2                    & 60           & 433               & 1.46            \\
2020-05-21 & 19:02:51   & 19:03:30 & 2       & 26               & 287                         & 42                         & 18.7       & 0.9                    & 42           & 276               & 1.24            \\
2020-05-21 & 19:03:31   & 19:04:05 & 1       & 20               & 53                          & 25                         & 20.9       & 3.2                    & 36           & 390               & 1.59            \\
2020-05-21 & 19:03:33   & 19:04:05 & 2       & 31               & 264                         & 49                         & 12.0       & 0.4                    & 34           & 255               & 1.26            \\
2021-09-13 & 00:13:55   & 00:14:45 & 1       & 23               & 54                          & 30                         & 17.5       & 1.8                    & 52           & 241               & 1.27            \\
2021-09-13 & 00:18:55   & 00:21:37 & 1       & 23               & 108                         & 19                         & 83.4       & 4.5                    & 164          & 387               & 1.46            \\
2021-09-13 & 00:14:17   & 00:17:07 & 2       & 19               & 68                          & 22                         & 156.6      & 32.2                   & 172          & 405               & 1.59            \\
2021-09-13 & 00:18:39   & 00:20:41 & 2       & 19               & 34                          & 9                          & 37.7       & 2.3                    & 124          & 323               & 1.41            \\
2021-09-13 & 00:18:39   & 00:19:01 & 2       & 19               & 53                          & 27                         & 19.5       & 5.2                    & 24           & 259               & 1.27            \\
\vdots          &            &          &         &                  & \vdots                           &                            &            &                        &              &                   & \vdots               \\
2023-04-11 & 23:40:16   & 23:41:31 & 3       & 21               & 77                          & 17                         & 35.7       & 2.2                    & 78           & 185               & 1.23            \\
2023-04-11 & 23:41:28   & 23:42:28 & 3       & 21               & 61                          & 21                         & 20.5       & 1.4                    & 63           & 185               & 1.14            \\
2023-04-11 & 23:38:52   & 23:41:31 & 4       & 17               & 30                          & 5                          & 152.7      & 19.9                   & 162          & 353               & 2.03            \\
2023-04-15 & 05:03:06   & 05:04:09 & 1       & 65               & 45                          & 24                         & 26.5       & 3.1                    & 66           & 185               & 1.19            \\
2023-04-15 & 05:06:03   & 05:10:36 & 1       & 65               & 115                         & 14                         & 255.4      & 32.3                   & 276          & 200               & 1.27            \\
2023-04-15 & 05:01:39   & 05:03:00 & 2       & 15               & 12                          & 7                          & 29.2       & 4.1                    & 84           & 179               & 1.56            \\
2023-04-15 & 04:53:27   & 04:57:24 & 4       & 7                & 36                          & 11                         & 105.4      & 7.5                    & 240          & 192               & 1.40            \\
2023-04-15 & 04:57:48   & 05:00:24 & 4       & 7                & 28                          & 8                          & 87.1       & 8.6                    & 159          & 220               & 1.52            \\ \hline
\end{tabular}
\end{adjustbox}
\tablefoot{The table gives the date and time, slit index, loop length, minor radius, displacement amplitude, uncertainty of the amplitude, period, uncertainty of the period, duration, minor radius, and intensity ratio. This table is available in its entirety\footnote{\url{https://github.com/DayeLim/OscillationParameter}}.}
\end{table*}

\section{Results and Discussion}\label{sec:result}

We find 147 oscillation events in about 71\% of QS loops (29 out of 41) and 107 oscillation events in about 86\% of AR loops (19 out of 22). This occurrence rate is consistent with that of about 70\% from coronal bright points in QS \citep{2022ApJ...930...55G} and that of about 90\% in ARs \citep{2015A&A...583A.136A}. In this study, we consider a duty cycle defined as duration divided by period to indicate how long oscillations last compared to period. The duty cycle of all oscillations ranges from 1 to 6, with an average value of about 1.5. Among 254 oscillations, 78 oscillations have a higher quality than 2. Given that the feature of decayless oscillations is no significant damping, one can expect that the quality of decayless oscillations in general would be high. However, our result shows that most decayless oscillations have a low-duty cycle. This may be an effect of the fact that, as already mentioned above, the oscillating loop is dynamic and its brightness is not maintained throughout the observation. There have been no quantitative studies that presented the duty cycle of decayless oscillations. However, through the figures showing oscillation examples, it can be seen that decayless oscillations with duty cycles ranging from approximately 1 to 7 \citep{2015A&A...583A.136A, 2022ApJ...930...55G, 2023ApJ...944....8L, 2023ApJ...946...36P, 2024A&A...685A..36S} and up to about 40 \citep{2022MNRAS.513.1834Z} were observed.

We investigate the intensity ratio ($R_{I}$) of oscillating loops to backgrounds. Using the minor radius of the loop of each oscillation, we define the boundary of the loop and calculate the average intensity ($I_{\text{i}}$) inside the loop. For the background, we consider the minimum intensity outside the loop ($I_{\text{e}}$). 
The intensity ratio ranges from 1.1 to 2.5. The average value of QS loops is about $1.6\pm0.02$, which is slightly higher than the average ratio ($1.4\pm0.02$) of AR loops. These values are overall lower than the intensity ratios previously reported. \citet{2022ApJ...930...55G} reported an average intensity ratio of 4.1 for coronal bright points in the QS observed by SDO/AIA. For AR loops, using AIA observations, the average value of 2.5 was estimated \citep{2019ApJ...884L..40A, 2023NatSR..1312963Z}. \citet{2003ApJ...598.1375A} showed a density contrast of about 3.3 (roughly 11 in an intensity contrast) on average of AR loops observed by Transition Region and Coronal Explorer. It is difficult to conclude any specific trend in intensity ratios in different regions because the method of measuring the loop intensity ratio is different for each study and the estimated value may be affected by different imaging instruments.

The assumption that all decayless oscillations detected in this study are standing transverse mode enables us to estimate the energy flux ($F\,[\text{W}\,\text{m}^{-2}]$) and energy ($E$ [J]) using the following formulae \citep{2014ApJ...795...18V}, 
\begin{equation}\label{eq:energyflux}
    F=\frac{1}{2}f\rho \left(\frac{2\pi A}{P}\right)^{2}\left(\frac{2L}{P}\right),
\end{equation}
\begin{equation}\label{eq:energy}
    E=(\pi R^{2}L)\frac{1}{2}\rho \left(\frac{2\pi A}{P}\right)^{2},
\end{equation}
where $R$, $L$, and $\rho$ are the minor radius, length, and plasma density of the oscillating loop respectively, $f$ is the filling factor, and $A$ and $P$ are the displacement amplitude and period of the oscillation. We use the parameters deduced from the observations for $R$, $L$, $A$, and $P$. Equations \ref{eq:energyflux} and \ref{eq:energy} are only valid for low filling factors up to 10\%, thus we use the constant $f$ of 10\% for all oscillating loops. By assuming that all oscillating loops have the same uniform temperature (1 MK) and gravitational acceleration, we consider stratified $\rho$. The same formula for $\rho$ described in Equation (3) in \citet{2023ApJ...952L..15L} is used.

\subsection{Statistical properties of oscillation parameters}
\begin{figure*}
  \resizebox{\hsize}{!}{\includegraphics{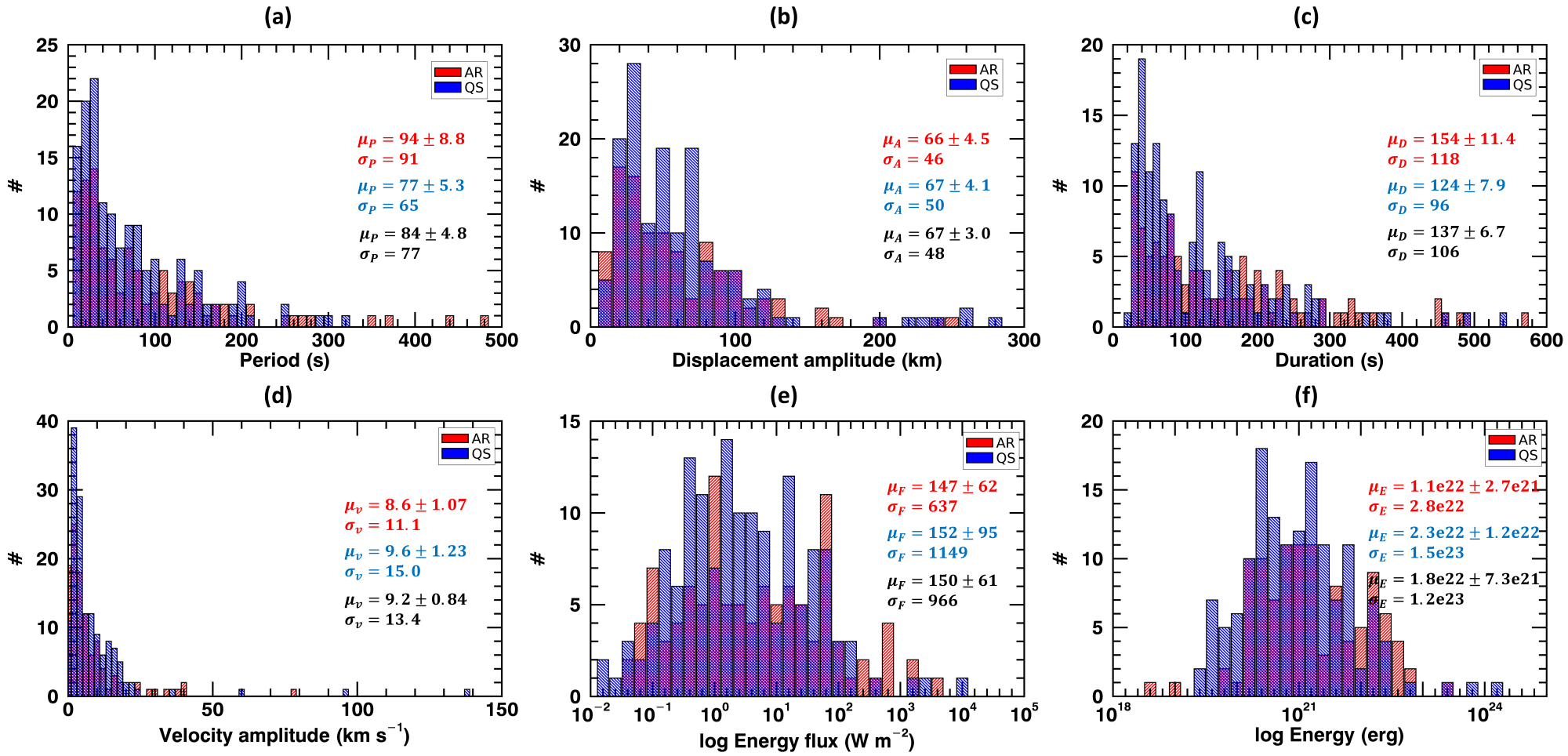}}
  \caption{Histogram of (a) the period, (b) the displacement amplitudes, (c) the duration, (d) the velocity amplitude, (e) the energy flux in log scale, and (f) the energy in log scale in quiet Sun (blue hatched bar) and active regions (red hatched bar). The average ($\mu$) and standard deviation ($\sigma$) of each parameter for quiet Sun (blue), active regions (red), and all regions (black) are indicated in each panel. Each average value is presented with its uncertainty as $\sigma/\sqrt{n}$ where $n$ is the number of detected oscillations. The physical unit of each parameter is shown on the $x$-axis in each panel.}
  \label{fig:hist_region}
\end{figure*}

The histograms of periods, displacement amplitudes, duration, velocity amplitudes, energy fluxes, and energies in each region are shown in Figure \ref{fig:hist_region}. 
The periods of all oscillations range from 12 to 490 s, with an average of $84\pm4.8$ s. This range falls within a period range between 11 s and 30 min of all detected decayless oscillations in the literature (see Figure 5 in \citealt{2023NatSR..1312963Z} and Figure 5 in \citealt{2024A&A...685A..36S}). For the QS we find periods between 12 and 323 s with an average of $77\pm5.3$. The periods of AR oscillations range from 15 to 490 s, with an average value of $94\pm8.8$ s. The period ranges of QS and AR are largely overlapped and AR loops tend to have longer periods on average. In our study, the average loop length ($72\pm11$ Mm) of AR loops is much longer than that ($27\pm3$ Mm) of QS loops. In the case of decayless oscillations following a trend of oscillations in \citet{2015A&A...583A.136A} (hereafter T1), the average periods for AR and QS are around 273 and 386 s respectively and both of their average loop lengths are around 232 Mm \citep{2013A&A...560A.107A, 2018ApJ...854L...5D, 2022MNRAS.513.1834Z}. In the case of decayless oscillations following a trend of oscillations in \citet{2022ApJ...930...55G} (hereafter T2), the average periods for AR and QS are around 49 and 202 s respectively and both of their average loop lengths are about 15 Mm \citep{2023ApJ...944....8L, 2023ApJ...946...36P, 2024A&A...685A..36S}. If we assume that all decayless oscillations are the fundamental harmonic and estimate their average phase speed ($2L/P$), the velocities of the oscillations in this study are about 701 $\text{km}\,\text{s}^{-1}$ for QS and 1532 $\text{km}\,\text{s}^{-1}$ for AR. The speeds of T1 oscillations are about 1202 $\text{km}\,\text{s}^{-1}$ for QS and 1700 $\text{km}\,\text{s}^{-1}$ for AR. For T2 oscillations, the estimated phase speeds are 149 $\text{km}\,\text{s}^{-1}$ for QS and 612 $\text{km}\,\text{s}^{-1}$ for AR. From this perspective, the oscillations detected in this study are likely between the two trends T1 and T2.      

The displacement amplitudes of QS oscillations are detected over a range from 12 to 287 km, which is similar to that of decayless oscillations previously observed in the QS, which have an amplitude range of 27 - 365 km \citep{2022ApJ...930...55G, 2024A&A...685A..36S}. The amplitudes of AR oscillations range from 11 to 252 km, which is within the amplitude range of previously reported AR decayless oscillations (24 - 500 km; \citealt{2015A&A...583A.136A, 2022A&A...666L...2M, 2022MNRAS.516.5989Z, 2023ApJ...944....8L}). There were no flares occurring between observation durations of AR sequences. Except for the case on 2023-03-29 that had 5 C-class flares and 1 X-class flare from about 12 hours to 5 hours before the observation start time, the rest did not have any activity at all even the day before. \citet{2021A&A...652L...3M} reported that solar flares enhanced displacement amplitudes compared to the pre-flare oscillations. In our case, although the time of the flares is not directly connected to the time when the oscillation is discovered, we compare the detected amplitudes in each AR region. It is shown that the amplitudes (11-162 km) in the flaring AR are smaller than those (15-252 km) in relatively stable ARs (observations on 2022-03-30 and 2023-04-07). This result may be an effect of different magnetic tensions. If we assume that the coronal magnetic field strengths in ARs are similar to each other, the difference in magnetic curvatures can determine the difference in magnetic tensions. The curvature (the reciprocal of the minor radius of the loop) in the stable ARs is roughly 2.5 times larger than the curvature in the flaring AR. This may result in higher amplitudes of the loops in the stable ARs.

We also find that there is no significant difference in amplitudes between QS and AR when compared to their average values ($67\pm4.1$ and $66\pm4.5$ km). Given the pixel plate scale (about 100-220 km) of the observations considered in this study, we detect sub-resolution displacement amplitudes. \citet{2022ApJ...930...55G} confirmed that SDO/AIA with a pixel plate scale of 435 km can observe a decayless oscillation with an amplitude of around 27 km using a 1D model. It could be expected that $\hrieuv$, which has up to about four times better spatial resolution than AIA, can detect oscillations with smaller amplitudes than observed by AIA. However, \citet{2022MNRAS.516.5989Z} reported on decayless oscillations observed simultaneously when AIA and $\hrieuv$ were aligned and showed that $\hrieuv$ tended to underestimate the amplitude. Thus, we can not exclude the influence of the instrument on the estimated oscillation parameters \citep{2024MNRAS.527.5302M}.

Figure \ref{fig:hist_region} (c) shows that the average duration ($154\pm11.4$ s) of AR oscillations is longer than that ($124\pm7.9$ s) of QS oscillations. The oscillation duration could be limited to the observational duration and 4 data sequences observing QS have a short observational duration (less than 1 min) compared to other sequences (up to one hour). When we excluded these four sequences, the average duration of QS oscillations was $127\pm8.1$ s, implying that the average value of all QS oscillations was not affected by the observational duration. Thus, we could interpret that decayless transverse oscillations in ARs could last longer than those in QS.

We investigate the velocity amplitude ($v=2\pi A/P$) in each region. This has a range between about 0.7 and 138 $\text{km}\,\text{s}^{-1}$ similar to the range in AR decayless oscillations \citep{2023ApJ...944....8L}, however, most oscillations have a velocity amplitude of less than around 20 $\text{km}\,\text{s}^{-1}$ as in \citet{2016A&A...591L...5N}, \citet{2022ApJ...930...55G}, and \citet{2024A&A...685A..36S}. We find that the average velocity amplitude of QS oscillations is rather higher than that of AR oscillations. This is because the displacement amplitude has a similar average value regardless of the region, whereas the period is longer in ARs.

The energy flux generated by QS oscillations ranges from about $2.0\times10^{-2}$ to $1.3\times10^{5}\,\text{W}\,\text{m}^{-2}$, with an average value of $152\pm95\,\text{W}\,\text{m}^{-2}$. This is broader than the energy flux range reported in \citet{2024A&A...685A..36S} ($0.6-314\,\text{W}\,\text{m}^{-2}$ in QS and CH) and \citet{2023ApJ...946...36P} (1900 and 6500 $\text{W}\,\text{m}^{-2}$ in QS). Among 147 oscillations, it is found that only 4 oscillations have the energy flux greater than 300 $\text{W}\,\text{m}^{-2}$, which is the energy loss corresponding to the QS \citep{1977ARA&A..15..363W}. The AR oscillations generate energy fluxes with a narrower range from about 0.06 to 6000 $\text{W}\,\text{m}^{-2}$, with an average value of $147\pm62\,\text{W}\,\text{m}^{-2}$. This average energy flux is much smaller than previously reported ($815\,\text{W}\,\text{m}^{-2}$) in AR loops \citep{2023ApJ...944....8L}. We do not find any single oscillations that can sufficiently heat the AR to counteract the energy loss ($\approx 10^{4}\,\text{W}\,\text{m}^{-2}$; \citealt{1977ARA&A..15..363W}). The average energy fluxes for each region in this study are comparable to each other. We would like to note that the energy fluxes estimated in \citet{2023ApJ...946...36P}, \citet{2023ApJ...944....8L}, and \citet{2024A&A...685A..36S} were the values when the filling factor was not considered. We will revisit the energy flux estimated in this study in a statistical view later on.

The range of the estimated energy and their average value for each region is shown in Figure \ref{fig:hist_region} (f). The energy of decayless oscillations ranges from about $10^{18}$ to $10^{24}$ erg, corresponding to femto, pico, and nanoflare energies, meaning that we have shown that there are decayless oscillations with lower energies than the decayless oscillations in the literature ($10^{20}-10^{24}$ erg; \citealt{2023ApJ...952L..15L}). Similar to the energy flux, there is no significant difference in the average energy between QS and AR. The omnipresence in both QS and ARs of decayless oscillations and their small energy characteristics are reminiscent of small-scale brightenings believed to be miniature solar flares, such as nanoflares \citep{1988ApJ...330..474P, 2013ApJ...771...21W, 2021A&A...647A.159C, 2022A&A...661A.149P}, extreme ultraviolet (EUV) transient brightenings \citep{1998A&A...336.1039B, 1999SoPh..186..207B}, campfires \citep{2021A&A...656L...4B, 2021A&A...656L...7C}, jets \citep{2023Sci...381..867C}, and explosive events \citep{2019ApJ...887...56T}. As suggested by \citet{2023ApJ...944....8L}, the decayless oscillations could be excited by small-scale flares as a consequence of the continuous motion of the footpoints in the photospheric convection. 
To confirm this conjecture, however, further study should be made.

\subsection{Correlation between oscillation parameters} 
\begin{figure*}
  \resizebox{\hsize}{!}{\includegraphics{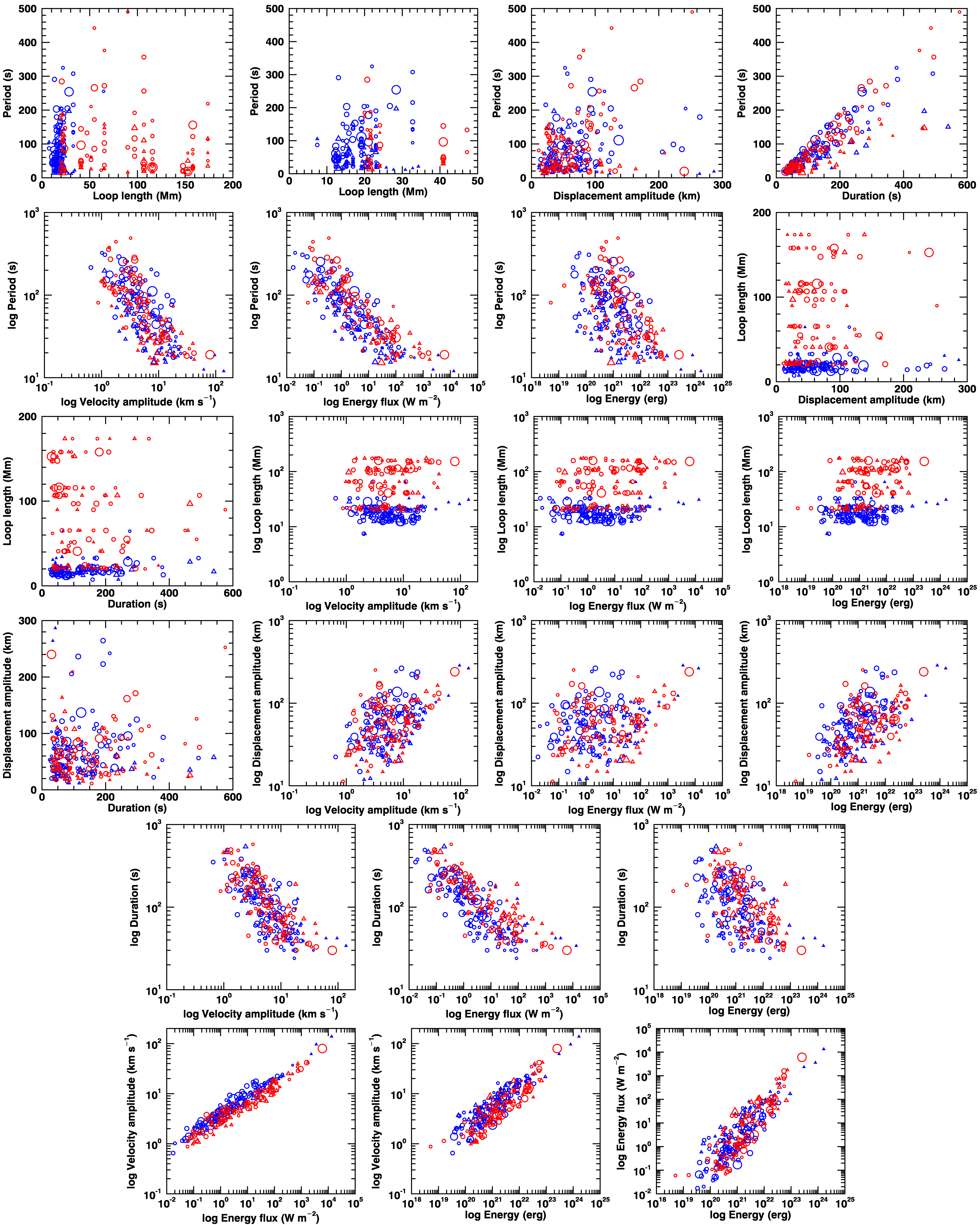}}
  \caption{Scatter plots between periods, loop lengths, displacement amplitudes, duration, velocity amplitudes, energy fluxes, and energies depending on coronal regions. The red and blue indicate active regions and quiet Sun respectively. A circle (triangle) represents a duty cycle of less (larger) than 2. The size of data points is proportional to the intensity ratio.}
  \label{fig:scatterplots}
\end{figure*}

We investigate the relationship between oscillation parameters in different coronal regions. Figure \ref{fig:scatterplots} shows the scatter plots and Table \ref{tab:cc} lists the correlation coefficients (CCs) between them. The CC between periods and loop lengths is $0.26\pm0.08$ for QS and $-0.14\pm0.10$ for AR, displaying a considerable difference. The CCs in each region indicate no correlation, which is consistent with the results for QS \citep{2022ApJ...930...55G, 2024A&A...685A..36S}, however, contrary for AR \citep{2015A&A...583A.136A, 2023ApJ...944....8L}. It is found that the CC between periods and amplitudes is $0.20\pm0.08$ for QS and $0.38\pm0.10$ for AR. \citet{2022ApJ...930...55G} and \citet{2024A&A...685A..36S} presented the CC of 0.4 and 0.52, respectively, between periods and amplitudes of QS and CH oscillations. \citet{2016A&A...591L...5N} showed that the amplitude gradually increases with the period using decayless oscillations analysed in \citet{2015A&A...583A.136A}, giving a CC of 0.67. We note that we calculate this CC value ourselves because it was not presented in \citet{2015A&A...583A.136A} and \citet{2016A&A...591L...5N}. The loop length and amplitude have no significant correlation (CC of about 0.1 for both QS and AR), which is consistent with decayless oscillations in QS \citep{2022ApJ...930...55G, 2024A&A...685A..36S} and even with decaying oscillations \citep{2019ApJS..241...31N}. 
Our result shows a negative correlation between periods and velocity amplitudes for both QS (-0.33) and AR (-0.39). \citet{2016A&A...591L...5N} showed that self-oscillations, which is one of the possible excitation mechanisms of decayless oscillations, can give consistent results with the relationship between periods and loop lengths (CC of 0.71) and between periods and velocity amplitudes (CC of -0.08) from the oscillations presented in \citet{2015A&A...583A.136A}. These two relationships in our study show the opposite tendency (however, the similar tendency to the results in \citealt{2022ApJ...930...55G} and \citealt{2024A&A...685A..36S}), implying that this mechanism may be the least favorable to the oscillations detected in this study. 

We find that the oscillation duration is highly correlated with the period but not the loop length and amplitude. This is perhaps partially caused by the fact that necessarily periods should be shorter than durations. Considering that periods, displacement amplitudes, and loop lengths do not show a significant correlation between them, we can expect the linear slope with the energy flux in log scales using Equation \ref{eq:energyflux} by assuming that they are independent of each other. However, a discrepancy between the expected values (-3 between periods-energy fluxes and 2 between amplitudes-energy fluxes) and the empirical results (-2.6 between periods-energy fluxes and 1.2 between amplitudes-energy fluxes) is seen. We also consider the relationship between the duty cycle, the intensity ratio, and oscillation parameters. They do not show a high correlation with oscillation parameters (absolute values less than 0.4) as shown in Figure \ref{fig:scatterplots}. In the case of the duty cycle, periods and displacement amplitudes have a negative correlation of around -0.3 and the CC of the other parameters are close to zero. Among oscillation parameters, periods and duration have a positive relationship, and loop lengths and velocity amplitudes have a negative relationship, showing a higher correlation (around $\pm0.4$) than other parameters.

\begin{figure*}
  \resizebox{\hsize}{!}{\includegraphics{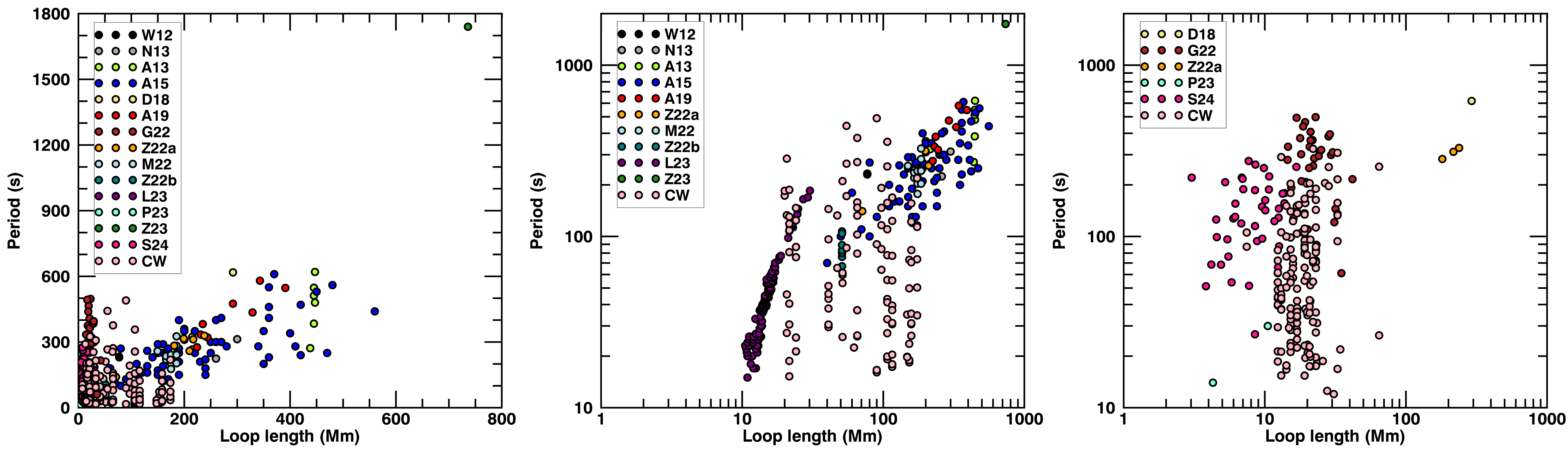}}
  \caption{Left panel: scatter plots between periods and loop lengths of decayless oscillations analysed in the current work (CW) and previous studies (from top to bottom in the legend, \citealt{2012ApJ...751L..27W, 2013A&A...552A..57N, 2013A&A...560A.107A, 2015A&A...583A.136A, 2018ApJ...854L...5D, 2019ApJ...884L..40A, 2022ApJ...930...55G, 2022MNRAS.513.1834Z, 2022A&A...666L...2M, 2022MNRAS.516.5989Z, 2023ApJ...944....8L, 2023ApJ...946...36P, 2023NatSR..1312963Z, 2024A&A...685A..36S}). The scatter plots for ARs (middle panel) and QS (right panel) are presented in log scales.}
  \label{fig:scatterplot_meta}
\end{figure*}

Figure \ref{fig:scatterplot_meta} presents the variation in the period with the loop length in the current study, combined with previous reports that provided the periods and loop lengths of decayless oscillations \citep{2012ApJ...751L..27W, 2013A&A...552A..57N, 2013A&A...560A.107A, 2015A&A...583A.136A, 2018ApJ...854L...5D, 2019ApJ...884L..40A, 2022ApJ...930...55G, 2022MNRAS.513.1834Z, 2022A&A...666L...2M, 2022MNRAS.516.5989Z, 2023ApJ...944....8L, 2023ApJ...946...36P, 2023NatSR..1312963Z, 2024A&A...685A..36S} in different regions. It could be seen that most oscillations are overlapped with other decayless oscillations previously reported. In our study, however, a new type of oscillation with a shorter period in longer loops (about 100-200 Mm) is also found. Consequently, the phase speed of these oscillations can reach up to $10^{4}\,\text{km}\,\text{s}^{-1}$, a few times higher than the typical Alfv\'{e}n speed in the corona. Most of the loops where these oscillations appear are long loops that are faintly visible above the noisy background. Therefore, we cannot rule out the possibility that these short-period oscillations occur not in the long loops considered, but in a shorter loop in the background or a structure in the lower atmospheric layer. If the loop lengths for these oscillations are not overestimated, an other possibility can be considered.  \citet{2023MNRAS.526..499H} performed a 1D numerical model describing a loop with a length of about 120 Mm and found that the contribution of the lower atmosphere to the harmonics is significant when the oscillations are driven by the photospheric motions, which corresponds to decayless oscillations. In this case, the transverse motions at the loop apex are not the fundamental harmonic but likely third harmonics. Then, by taking this into account, the phase speed is one-third of the original value and this is comparable to the coronal Alfv\'{e}n speed. Note that we do not exclude other possibilities and interpretations and to more clearly understand this, more observational and theoretical information would be needed.

\subsection{The distribution of the energy flux and oscillation frequency} 
Using the estimated energy flux, the logarithm of the energy flux of each oscillation with a constant frequency bin (size of 0.1 in log scale) is considered. We note that \citet{2023ApJ...952L..15L} estimated the total energy flux per frequency bin. In tracking the center position of a loop, even though it is a single loop, if it is interrupted by another loop or background during observation, the center of the loop is hardly continuously detected. As a result, one oscillation may be returned as multiple oscillations with similar periods. In this case, the sum of energy flux or energy from each oscillation could be overestimated. However, it is expected that even if multiple oscillations are detected for one actual oscillation, it will not affect the average value. Thus, in this study, we consider the average energy flux for each frequency bin. 

\begin{figure}
  \resizebox{\hsize}{!}{\includegraphics{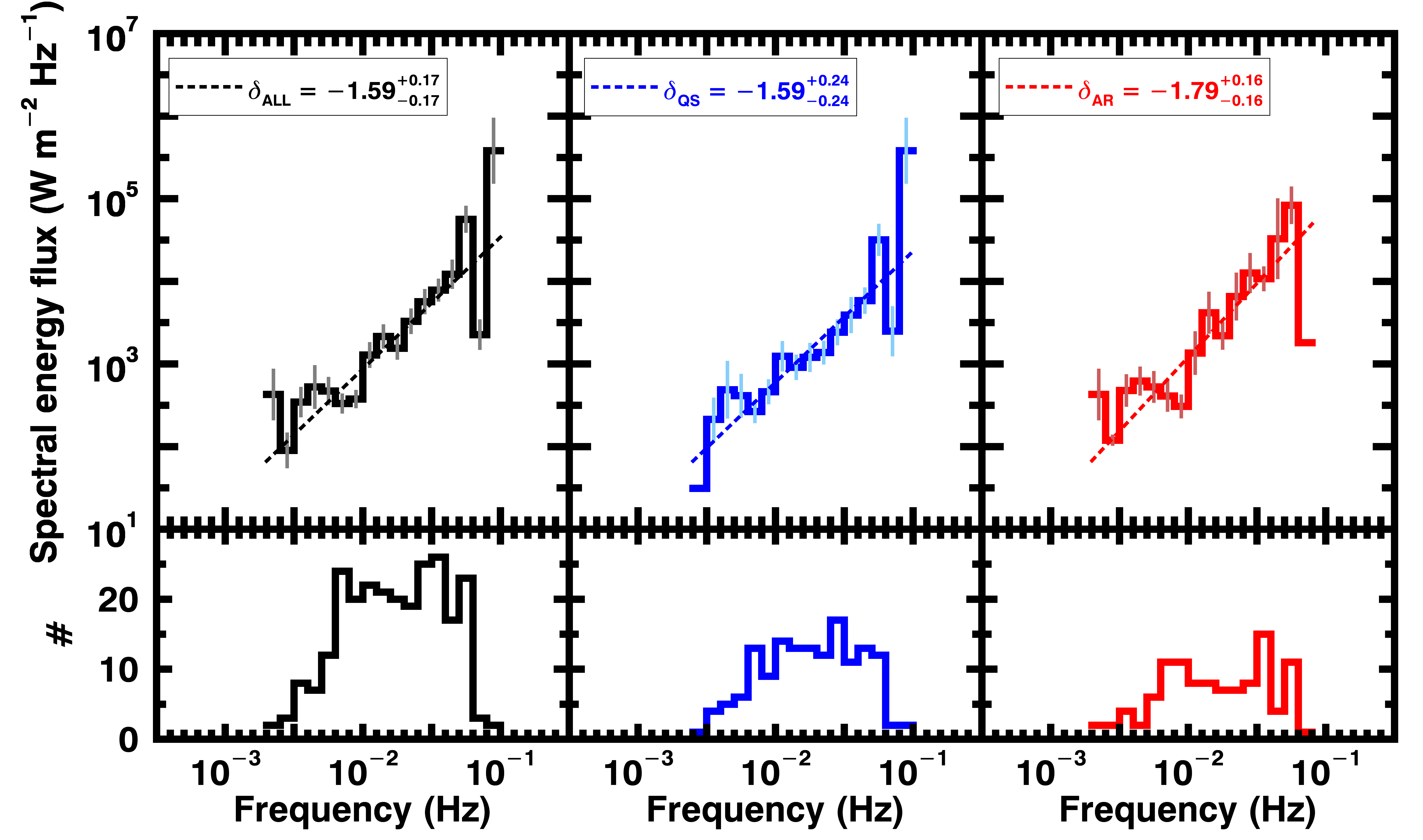}}
  \caption{The distribution, $s(\omega)$, of spectral energy fluxes as a function of oscillation frequencies (top panels) and the number of oscillations for each frequency bin (bottom panels). The vertical bars show an uncertainty ($\sigma_{\omega}=\sigma_{Fi}/\sqrt{n_{\omega i}}$). A bin size of 0.1 has been considered. The best fits of distributions are shown in dashed lines. The fitting was only considered for bins with a number of cases greater than 1. Blue, red, and black represent oscillations in the quiet Sun, active regions, and all regions, respectively. The power law slopes are $\delta_{\text{QS}}=-1.59\pm0.24$, $\delta_{\text{AR}}=-1.79\pm0.16$, and $\delta_{\text{ALL}}=-1.59\pm0.17$, respectively.}
  \label{fig:distribution}
\end{figure}

The estimated average spectral energy flux is shown in Figure \ref{fig:distribution}. The distribution has an uncertainty corresponding to the standard deviation of energy fluxes for each bin divided by the square root of the number of oscillations per bin. In order to estimate the best power-law fit of the distribution and its credible interval, we use the Solar Bayesian Analysis Toolkit (SoBAT; \citealt{2021ApJS..252...11A}). The logarithmic uncertainties of each bin are taken into account in the fit. The fitting was only considered for bins with a number in cases greater than 1. It is shown that there is a difference between the distributions in QS and AR. The power-law slope ($\delta$) of spectral energy flux from decayless oscillations observed in QS is around $\delta_{\text{QS}}=-1.59\pm0.24$ between frequency bins of about 0.003 and 0.09 Hz. In the case of AR oscillations with frequencies ranging from about 0.002 to 0.07 Hz, the slope of spectral energy flux is around $\delta_{\text{AR}}=-1.79\pm0.16$. This result that the slope is steeper in AR than in QS is consistent with the result in \citet{2016ApJ...828...89M}. When we consider all observed decayless oscillations, a general power-law form has a slope of $\delta_{\text{ALL}}=-1.59\pm0.17$ between the frequency of about 0.002 and 0.09 Hz. Similar to the spectral slope of -1.4 obtained from previously reported decayless oscillations \citep{2023ApJ...952L..15L}, it is clear that regardless of coronal regions, the slopes are all much less than the critical slope of $\delta = 1$. This implies that high-frequency oscillations statistically contribute more to heating than low-frequency oscillations in QS and ARs, respectively.

\begin{figure}
  \resizebox{\hsize}{!}{\includegraphics{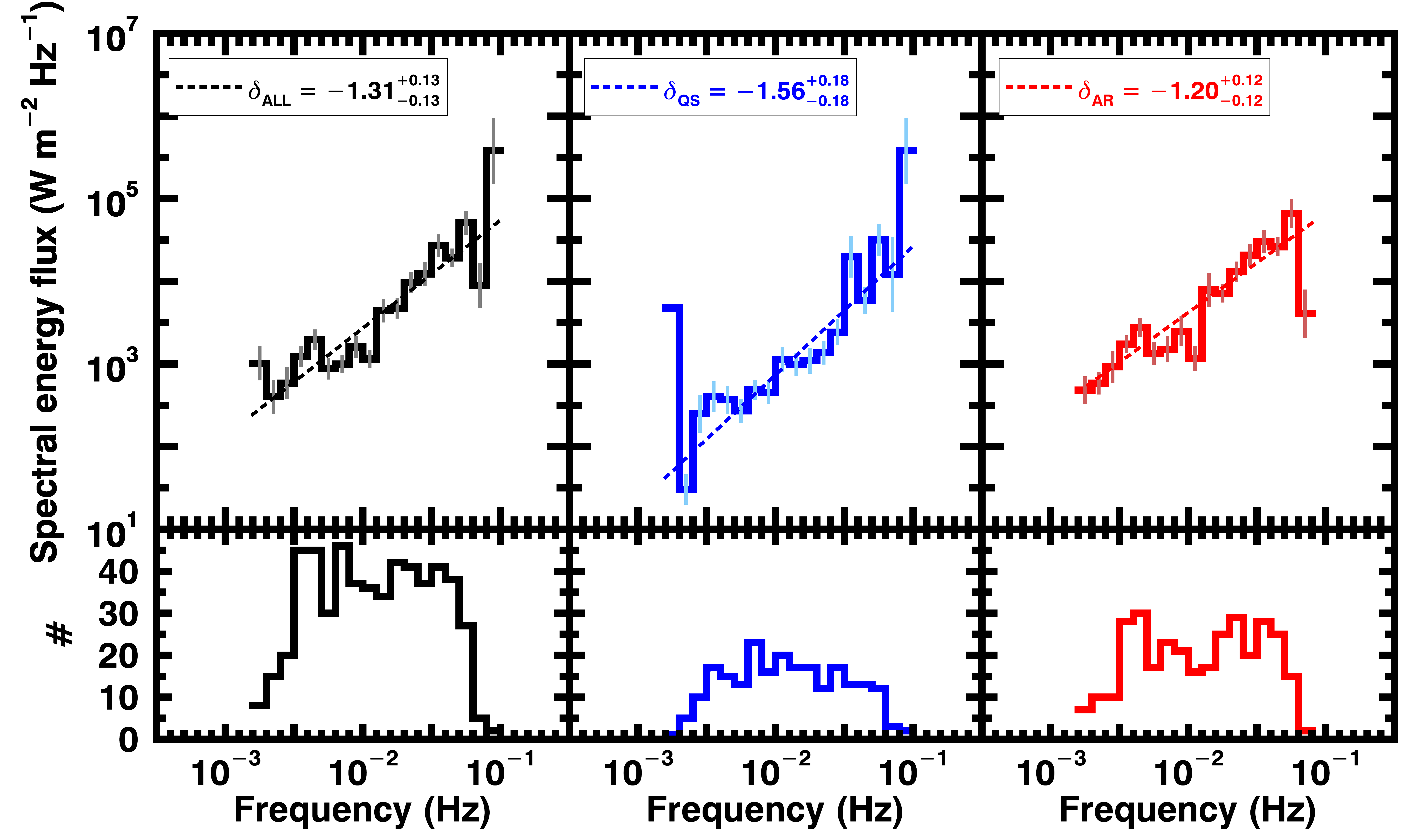}}
  \caption{Same as Figure \ref{fig:distribution} but for the oscillations including the decayless oscillations in this study and previous studies \citep{2013A\string&A...560A.107A, 2015A\string&A...583A.136A, 2018ApJ...854L...5D, 2022ApJ...930...55G, 2022MNRAS.513.1834Z, 2023ApJ...946...36P, 2022MNRAS.516.5989Z, 2022A\string&A...666L...2M, 2023ApJ...944....8L, 2024A&A...685A..36S}. The oscillations from previous studies are the same listed in Table 1 in \citet{2023ApJ...952L..15L}. The power law slopes are $\delta_{\text{QS}}=-1.56\pm0.18$, $\delta_{\text{AR}}=-1.20\pm0.12$, and $\delta_{\text{ALL}}=-1.31\pm0.13$, respectively.}
  \label{fig:distribution_meta}
\end{figure}

Figure \ref{fig:distribution_meta} shows the average spectral energy flux of decayless oscillations in our work combined with various studies. The list of studies considered for this combined distribution is presented in Table 1 in \citet{2023ApJ...952L..15L}. Contrary to the result of the oscillations analysed in this study, we can see that the spectral slope in QS is steeper than that in ARs. However, their slopes are still less than the critical slope, emphasizing the significant role of high-frequency oscillations (near 0.09 Hz for QS and 0.07 Hz for ARs). We provide an assessment of the potential amount of average energy flux that could be dissipated in the different coronal regions using Equation \ref{eq:approx_energyflux_high}. From the empirical fitting parameters of the power-law in QS ($s_{0}\approx9.5\times10^{5}$ and $\omega_{\text{max}}\approx0.09\,\text{Hz}$), the estimated total average energy flux carried by QS decayless oscillations is about 785 $\text{W}\,\text{m}^{-2}$. Given the energy loss of roughly 300 $\text{W}\,\text{m}^{-2}$ in the QS \citep{1977ARA&A..15..363W, 2006SoPh..234...41K}, this value indicates that the QS could be sufficiently heated by high-frequency decayless oscillation. In the case of ARs ($s_{0}\approx1.1\times10^{6}$ and $\omega_{\text{max}}\approx0.07\,\text{Hz}$), the value is about 1400 $\text{W}\,\text{m}^{-2}$, which is less than the heating requirement for ARs (about $10^{4}$ $\text{W}\,\text{m}^{-2}$). If we assume that $s_{0}$ and $\delta$ are constant, we can speculate the required minimum period for balancing the energy loss in ARs. We find this to be 6 s (0.17 Hz), which is ideally observable from $\hrieuv$ when they target a cadence of 1 s. In this study, one data set with a 1-s cadence was included but no oscillations were detected in there. No conclusions about whether there are higher frequency oscillations or not can be drawn based on only one case because the observational duration of this data set was only 18 s. Thus more high cadence $\hrieuv$ observations will be needed to clarify this. 

We would like to note that the energy fluxes considered in this study are calculated assuming that the oscillating loops are very dense, i.e., external densities are ignored. However, based on the result of this study with other observational results, it seems that the plasma density between the inside and outside of the loop is not significantly different in the solar corona. By taking this into account, the estimated energy fluxes increase by a factor of 1.8. Moreover, the estimated energy flux does not correspond to the amount it dissipates. \citet{2020ApJ...897L..13H} presented an analytical model of the dissipation rate by assuming the KHi-induced steady-state turbulence. Using the Equation (6) in \citet{2020ApJ...897L..13H}, we find the dissipation rate ranging from about 1 to 43\%. If we consider the factor of 1.8 and these minimum and maximum dissipation rates for each oscillation, we approximate the dissipated energy fluxes of around 14 and 608 $\text{W}\,\text{m}^{-2}$ in the QS and 25 and 1080 $\text{W}\,\text{m}^{-2}$ in the ARs. In the case of ARs, the result remains unchanged that decayless oscillations can not sufficiently heat the AR statistically, but in the QS, this result can be seen to depend on the dissipation rate. 
These results will be updated when more precise theoretical and numerical studies on the dissipation rate of decayless oscillations are performed. 

\begin{table*}
\caption{The linear Pearson correlation coefficients (CCs) between periods ($P$), loop lengths ($L$), displacement amplitudes ($A$), duration ($D$), velocity amplitudes ($v$), energy fluxes ($F$), and energies ($E$) depending on coronal regions. $\sigma$ is the standard error ($1/\sqrt{n}$), where $n$ is the number of oscillations. The logarithmic scale is considered when calculating the correlation coefficient with the energy flux or energy.}
\label{tab:cc}
\begin{adjustbox}{width=1\textwidth}
\small
\begin{tabular}{ccccccccccccccccccccccc}
\hline
    & $\sigma$ & $\text{CC}_{P-L}$ & $\text{CC}_{P-A}$ & $\text{CC}_{P-D}$ & $\text{CC}_{P-v}$ & $\text{CC}_{P-F}$ & $\text{CC}_{P-E}$ & $\text{CC}_{L-A}$ & $\text{CC}_{L-D}$ & $\text{CC}_{L-v}$ & $\text{CC}_{L-F}$ & $\text{CC}_{L-E}$ & $\text{CC}_{A-D}$ & $\text{CC}_{A-v}$ & $\text{CC}_{A-F}$ & $\text{CC}_{A-E}$ & $\text{CC}_{D-v}$ & $\text{CC}_{D-F}$ & $\text{CC}_{D-E}$ & $\text{CC}_{v-F}$ & $\text{CC}_{v-E}$ & $\text{CC}_{F-E}$ \\ \hline \hline
AR  & 0.10  & -0.14  & 0.38   & 0.88   & -0.39  & -0.86            & -0.60            & 0.12   & -0.15  & 0.24   & 0.35             & 0.37             & 0.18   & 0.52   & 0.27             & 0.52             & -0.45  & -0.83            & -0.61            & 0.97   & 0.90   & 0.88             \\
QS  & 0.08  & 0.26   & 0.20   & 0.86   & -0.33  & -0.85            & -0.55            & 0.14   & 0.22   & 0.16   & 0.09             & 0.22             & 0.05   & 0.56   & 0.31             & 0.59             & -0.33  & -0.80            & -0.52            & 0.97   & 0.91   & 0.87             \\
ALL & 0.06  & 0.02   & 0.28   & 0.87   & -0.34  & -0.84            & -0.56            & 0.06   & 0.04   & 0.08   & 0.22             & 0.28             & 0.11   & 0.55   & 0.29             & 0.55             & -0.37  & -0.79            & -0.53            & 0.96   & 0.89   & 0.88             \\ \hline
\end{tabular}
\end{adjustbox}
\end{table*}

\section{Conclusions}\label{sec:conclusion}
Here, we have analysed decayless oscillations in the QS and ARs observed by the SolO/EUI $\hrieuv$. We have identified 147 oscillations in 41 QS loops and 107 oscillations in 22 AR loops. The range of detected periods, displacement amplitudes, and loop lengths were comparable to that of decayless oscillations in previous reports. Our statistical study depending on coronal regions reveals that periods and loop lengths show a significant difference in different regions but amplitudes do not. We found no linear correlation between periods and loop lengths in the QS and even in ARs. A few decayless oscillations in the QS were found to have an energy flux of greater than about 300 $\text{W}\,\text{m}^{-2}$ equivalent to the energy loss in the QS. However, in AR, there was no oscillation with an energy flux that could compensate for the energy loss. The estimated energy of decayless oscillations corresponded to the range of femtoflare and picoflare. The distribution of spectral energy fluxes was described as a power-law, showing a steeper slope in ARs than in the QS. It was found that both slopes are less than the critical slope of 1, implying that high-frequency decayless oscillations could play a key role in coronal heating. The total average energy flux estimated from the empirical fitting showed that the QS could be sufficiently heated by decayless oscillations, however, they cannot support the heating in AR, unless yet-unobserved oscillations are present with frequencies up to 0.17Hz.

Our results indicate the statistical importance of high-frequency oscillations in both QS and AR. The discovery of high-frequency oscillations depends on the temporal cadence of the instrument. Future EUI campaigns with high cadence observations ($\leq 1$ s)  will allow us to make further analyses of the energy flux distribution and its spectral slope in the future.

\begin{acknowledgements}
      Solar Orbiter is a space mission of international collaboration between ESA and NASA, operated by ESA. The EUI instrument was 
      built by CSL, IAS, MPS, MSSL/UCL, PMOD/WRC, ROB, LCF/IO with funding from the Belgian Federal Science Policy Office (BELSPO/PRODEX PEA C4000134088); the Centre National d’Etudes Spatiales (CNES); the UK Space Agency (UKSA); the Bundesministerium für Wirtschaft und Energie (BMWi) through the Deutsches Zentrum für Luft- und Raumfahrt (DLR); and the Swiss Space Office (SSO). DL was supported by a Senior Research Project (G088021N) of the FWO Vlaanderen. TVD was supported by the European Research Council (ERC) under the European Union's Horizon 2020 research and innovation programme (grant agreement No 724326), the C1 grant TRACEspace of Internal Funds KU Leuven, and a Senior Research Project (G088021N) of the FWO Vlaanderen. The research benefitted greatly from discussions at ISSI. Furthermore, TVD received financial support from the Flemish Government under the long-term structural Methusalem funding program, project SOUL: Stellar evolution in full glory, grant METH/24/012 at KU Leuven. The research that led to these results was subsidised by the Belgian Federal Science Policy Office through the contract B2/223/P1/CLOSE-UP. This project DynaSun has received funding under the Horizon Europe programme of the European Union under grant agreement (no. 101131534). Views and opinions expressed are however those of the author(s) only and do not necessarily reflect those of the European Union and therefore the European Union cannot be held responsible for them.
\end{acknowledgements}

\bibliographystyle{aa} 
\bibliography{Lim_bib} 

\end{document}